\documentclass[AMA,STIX1COL]{WileyNJD-v2}
\usepackage{moreverb}

\newcommand\BibTeX{{\rmfamily B\kern-.05em \textsc{i\kern-.025em b}\kern-.08em
T\kern-.1667em\lower.7ex\hbox{E}\kern-.125emX}}

\usepackage{amsmath,amssymb,amsfonts,bm}
\usepackage{etoolbox}
\apptocmd{\thebibliography}{\setlength{\itemsep}{0pt}}{}{}

\DeclareMathOperator*{\argmax}{arg\,max}

\articletype{Article Type}%

\received{<day> <Month>, <year>}
\revised{<day> <Month>, <year>}
\accepted{<day> <Month>, <year>}

\begin{document}

\title{Distributed Pricing-based Resource Allocation for Dense D2D Communications in Beyond 5G Networks}

\author[1]{Mohammad Hossein Bahonar}

\author[1]{Mohammad Javad Omidi}

\authormark{BAHONAR \textsc{et al}}

\address[1]{Department of Electrical and Computer Engineering,
Isfahan University of Technology, Isfahan 84156-83111, Iran}

\corres{Mohammad Hossein Bahonar, 
Department of Electrical
and Computer Engineering, Isfahan University of Technology, Isfahan, 84156-
83111, Iran.\\
\email{mh.bahonar@ec.iut.ac.ir}}

\abstract[Abstract]{
Considering the dramatic increase of data rate demand in beyond 5G networks due to the numerous transmitting nodes, dense device-to-device (D2D) communications where multiple D2D pairs can simultaneously reuse a cellular link (CL) can be considered as a communication paradigm for future wireless networks.
Since distributed methods can be more practical compared to complex centralized schemes, we propose a low-complexity distributed pricing-based resource allocation algorithm to allocate power to cellular user equipments (CUEs) and D2D pairs {\color{black} constrained to the minimum quality-of-service} (QoS) requirements of CUEs and D2D pairs in two phases.
The price of each link is set by the base station (BS).
CUEs and D2D pairs maximize their rate-and-interference-dependent utility function by adjusting their power in the first phase, while the price of each link is updated at the BS based on the minimum QoS requirements of CUEs and D2D pairs in the second phase.
The proposed utility function controls D2D admission, hence it is suitable for dense scenarios.
{\color{black}
The proposed method is fast due to the closed-form solution for the first phase power allocation.}
Numerical results verify the effectiveness of the proposed method for practical dense beyond 5G scenarios by allocating resources to multiple D2D pairs and taking advantage of the spatial reuse gain of D2D pairs.
Furthermore, the utility function definition is discussed to illustrate its effectiveness for SE maximization.
Finally, the SE of the proposed method is compared to other centralized and distributed algorithms to demonstrate the higher sum-rate performance of the proposed algorithm.
}


\jnlcitation{\cname{%
\author{Bahonar M. H.}, 
\author{Omidi M. J},
\author{Yanikomeroglu H.}
} (\cyear{2016}), 
\ctitle{Distributed Pricing-based Resource Allocation for Dense D2D Communications}, 
\cjournal{Trans Emerging Tel Tech}, \cvol{2020;eXXXX}.
\url{https://doi.org/10.1002/ett.XXXX}}

\maketitle

\section{Introduction}\label{SecIntro}

With the development of new wireless generation systems such as 5th generation (5G) and 6th generation (6G), smart devices and new applications such as augmented reality (AR), virtual reality (VR), and vehicular Internet-of-Things (IoT) has emerged.\cite{S030}
Considering the new applications and the dense deployment of transmitting nodes in beyond 5G networks,\cite{S034} it is estimated that the data rate demand growth from 2010 to 2020 is more than 1000x and the number of connected devices in 2020 is almost 50 billion devices. \cite{S032,S033}

To achieve a high data rate, new communication paradigms such as massive multiple-input multiple-output (M-MIMO), reflecting intelligent surface (RIS), and device-to-device (D2D) communication should be developed to improve spectral efficiency (SE) of cellular networks.
Considering the dense deployment of transmitting nodes and to address the data rate demand challenge of beyond 5G cellular networks, D2D communications that enable direct data transmission among user equipments (UEs) by exploiting physical proximity of UEs can be a key technology for 5G new radio (NR) networks.\cite{S035,S036}
It has been shown that D2D communications can improve the overall SE of cellular networks by allowing D2D pairs to reuse the cellular spectrum and managing the interference properly {\color{black} while guaranteeing the minimum quality-of-service (QoS) requirements of cellular UEs (CUEs), D2D pairs, or both. \cite{S037,S038,R003,R016}}
Data offloading, energy efficiency (EE) improvement,\cite{S039} coverage extension,\cite{R144,R053} and battery life extension\cite{R024} are some of the other advantages of D2D communications.
{\color{black} 
Different tools such as game theory \cite{R009,S051}, stochastic geometry \cite{S039,S041}, convex optimization \cite{R003}, and machine learning \cite{S061,S062} have been used for resource allocation to D2D communications in the literature.
}

Outband operation corresponding to D2D communications in unlicensed bands has limited usage since it is not controlled by the base station (BS) and the minimum QoS requirements of the D2D pairs may not be guaranteed.\cite{R017}
Therefore, we study resource allocation for D2D communications in a licensed spectrum termed as an inband operation that can take place in an overlay or an underlay manner. \cite{R052}
Interference mitigation among D2D tier and cellular tier is one of the key challenges of D2D communications.
{\color{black} 
Dedicated resources are assigned to D2D pairs in the overlay inband scheme as it has been discussed in Reference \cite{S040} where the authors propose a three-phase overlay spectrum sharing scheme, hence the inter-tier interference issue is avoided.}
The study of the overlay inband scheme is less challenging and results in lower SE improvement compared to the study of the underlay inband scheme where D2D pairs can reuse the cellular spectrum of CUEs.
Reuse of the cellular spectrum is investigated and proposed by the majority of the literature on D2D communications.\cite{S041,R003,R049,S042}
{\color{black} 
The authors of Reference \cite{S042}, simultaneously address the interference management and energy consumption problems by efficiently allocating resources in energy harvesting-assisted underlay D2D communications.}

The resource allocation to D2D pairs and CUEs can be performed in a centralized  \cite{R003,R042_026} or a distributed manner \cite{R011}.
In Reference \cite{R003}, optimal resource allocation to D2D pairs using a centralized scheme is investigated but it is assumed that each cellular link (CL) can be used by at most one D2D pair, hence the investigated system model is not a dense scenario.
In References \cite{R042_026,R077}, it is assumed that multiple D2D pairs can reuse each CL which results in higher SE improvement compared to the case where at most one D2D pair can reuse each CL.
A heuristic approach is presented at Reference \cite{R077} to maximize the sum-rate of the cell.
The authors of Reference \cite{R042_026}, propose a subchannel sharing protocol followed by a greedy D2D selection procedure which can be used for dense scenarios but has high computational complexity.
The centralized approaches usually result in higher SE with higher computational complexity and signaling overhead compared to distributed schemes but are usually impractical for dense scenarios and beyond 5G networks.
The D2D pairs can reuse the downlink \cite{R014_031,R024_030,R046} or the uplink \cite{R005,R142,R025} cellular spectrum in an underlay inband scheme.
The reuse of the downlink spectrum results in lower SE improvement compared to the reuse of the uplink spectrum as the uplink spectrum is less utilized, and hence a larger number of D2D pairs can reuse the uplink spectrum due to its spatial reuse gain.\cite{R003_009,R077}
Therefore, we study distributed resource allocation for dense D2D communications undelaying the uplink spectrum of cellular networks.

Due to the dense deployment of transmitting nodes and D2D pairs in beyond 5G networks, distributed resource allocation methods can be considered as practical algorithms due to their low computational complexity.
Most of the distributed resource allocation methods use game-theoretic approaches to develop and propose efficient algorithms. \cite{R077_2,R088,R075,R042_026_018}
In Reference \cite{R077_2}, the spectrum trading problem has been investigated but one D2D pair is considered.
In Reference \cite{R088}, a power control mechanism has been proposed for D2D communications using game theory but one CUE is considered.
The authors of Reference \cite{R075} investigate cognitive D2D communications underlaying cellular networks using a double-sided bandwidth-auction game but one D2D pair is assumed to exist in the cell.
In Reference \cite{R042_026_018}, a distributed resource sharing algorithm for D2D communications which jointly considers mode selection, resource allocation, and power control is proposed.
However, it is assumed that each CL can be shared with at most one D2D pair and each D2D pair can reuse multiple CLs.
Additionally, the QoS constraints of D2D pairs are not considered.

Considering dense networks \cite{R138}, it is necessary to assume that the number of D2D pairs is larger than the number of CUEs.
In References \cite{R011,R042_026_018}, it is assumed that each D2D pair may reuse the CL of multiple CUEs simultaneously which is not a practical assumption from the hardware point of view since it is logical to assume that the transmitter of each D2D pair has one transmitter module.
The authors of Reference \cite{R011}, propose a pricing-based interference management algorithm.
However, the proposed method does not maximize the overall sum-rate of the cell but tries to guarantee the QoS constraints of CUEs using an interference management scheme.
Some other researches such as Reference \cite{R016,R042_026_025,R044,R043,R094} assume that each D2D pair can reuse at most one CL but multiple D2D pairs can reuse one CL which is a reasonable assumption from the hardware point of view but makes the resource allocation problem more challenging since binary resource sharing indicator variables may be needed in the resource allocation problem formulation to indicate resource sharing between D2D pairs and CUEs.

The authors of Reference \cite{R016}, propose an auction-based distributed resource allocation algorithm. 
However, the downlink spectrum of the cellular network is proposed to be shared with D2D pairs and it is assumed that the transmission power of each transmitter is selected from a finite set of transmission powers.
It is also assumed that the resource allocation to all D2D pairs is possible and the resource allocation problem is feasible considering all D2D pairs.
This assumption is not valid for dense scenarios since some of the D2D pairs may not be able to share resources with CUEs and may remain unadmitted.
Coalitional games have been used in Reference \cite{R042_026_025} for resource allocation of D2D communications in uplink cellular networks.
However, the minimum QoS requirements of CUEs and D2D pairs are not considered and the transmission powers are assumed to be fixed.
Joint mode selection and resource allocation of D2D communications has been investigated in Reference \cite{R044} using a matching game approach but transmission powers are assumed to be constant and are not optimized.
In Reference \cite{R043}, a Stackelberg game with pricing has been proposed to allocate resources for D2D communications in a time-varying environment.
However, the downlink spectrum is considered, the BS to CUE transmission power is assumed to be constant, and the proposed payoff function is a function of expected interference.
The proposed payoff function is not rate-dependent, and hence it may not be able to maximize the SE of the cell.
The authors of Reference \cite{R094} use coalitional games to allocate resources to D2D communications underlaying uplink cellular networks.
However, the transmission powers are assumed to be fixed and a selection procedure is utilized to maximize the sum-rate of the cell.

In this paper, we investigate D2D communications underlaying cellular networks considering the dense deployment of transmitting nodes in beyond 5G networks for uplink resource allocation.
{\color{black}
All CUEs which are primary users must remain admitted since the cell is fully-loaded. 
Hence, we aim to allocate transmission power to each CUE.
Considering the dense deployment of D2D pairs, the admission of each D2D pair may not be guaranteed. 
Therefore, we aim to allocate CL and transmission power to each D2D pair.
We propose a resource allocation algorithm in a distributed manner using a pricing-based approach due to its lower computational complexity compared to centralized schemes.}
By introducing price values for the interference links and an innovative utility function, we propose a novel two phases distributed pricing-based non-cooperative game for resource allocation where CUEs and D2D pairs are the players of the game.
In the first phase, the transmission powers of CUEs and the transmission powers and CLs of D2D pairs are adjusted in a distributed manner where each player of the game maximizes its utility function.
Considering the innovative proposed utility function, closed-form solutions are derived for transmission power adjustment problems, and hence the proposed scheme has low computational complexity.
In the second phase, the BS updates the prices of interference links using the efficient proposed updating scheme in order to guarantee the minimum QoS requirements of CUEs and D2D pairs.
We also develop fast and effective admission criteria for the admission of D2D pairs based on the proposed utility function.
Additionally, the signaling overhead is reduced significantly due to the innovative distance based definition of utility function.
Numerical results verify the effectiveness of the proposed resource allocation method for practical dense scenarios of beyond 5G networks.
The main contributions of the paper can be summarized as follows:
\begin{itemize}
\item
We consider dense deployment of D2D communications underlaying cellular networks where multiple D2D pairs can reuse a CL and each D2D pair may reuse at most one CL.
By formulating the SE maximization problem constrained to minimum QoS requirements of CUEs and D2D, we jointly allocate transmission power and spectrum to CUEs and D2D pairs in a decentralized manner that has not been investigated thoroughly in the literature.
\item
An innovative utility function that considers the data rate as a reward and the interference as a cost has been proposed which results in effectively distributed maximization of the overall-sum rate of the cell guaranteeing the minimum QoS requirements of CUEs and D2D pairs.
\item
Considering the locations of UEs, the innovative proposed utility function uses the distances between UEs to estimate the interference cost.
Therefore, the proposed algorithm has a low signaling overhead that is practical for dense scenarios.
\item
The transmission power adjustment of each CUE and the transmission power adjustment and CL selection of each D2D pair are modeled as convex and mixed-integer optimization problems, respectively.
The derived closed-form solutions cause the proposed algorithm to have low computational complexity.
\item
Considering the proposed utility function, a fast and effective D2D pair admission control has been proposed that causes the proposed method to be practical for dense scenarios.
\end{itemize}

The rest of the paper is organized as follows.
We describe the system model and formulate the resource allocation problem in Section \ref{SysModelSec}.
The phases of the distributed resource allocation problem are presented in Section \ref{PopAlgSec}.
In Section \ref{RemarkSec}, the proposed algorithm and some remarks are discussed. 
The simulation results are reported in Section \ref{SimulationSec}.
Section \ref{ConcSec} concludes the paper.

\section{System Model and Problem Formulation}
\label{SysModelSec}
In this section, we present the system model, formulate the interference and data rate of CUEs and D2D pairs, and then express the formulation of the resource allocation problem.

\subsection{System Model}
\begin{figure}[htbp]
    \centering
    \includegraphics[trim={0cm 0 0cm 0},clip,width=0.6\linewidth]{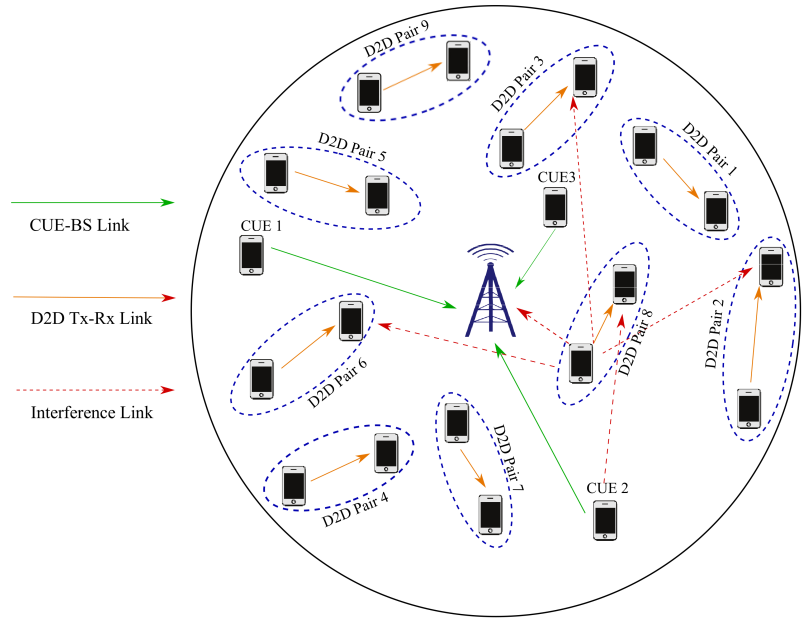}
    \caption{The system model of D2D communications underlaying a single cell dense cellular network.}
    \label{FigSysModel}
\end{figure}

We consider a single cell dense cellular network with $N$ orthogonal and equal bandwidth CLs as shown in Fig. \ref{FigSysModel}.
It is reasonable to assume a single cell network as interference management is the main challenge of D2D communications and can be managed efficiently \cite{R042_026_023_017}.
The BS which is located at the center of the cell has greater transmission power capability compared to CUEs and D2D pairs.
The downlink signal of the BS exists at most parts of the cell coverage area while the uplink signal of each CUE exists in its proximity.
Hence, the downlink spectrum has lower reuse capability compared to the uplink spectrum.
We consider the uplink spectrum due to its underutilization \cite{R003_009} and the fact that it can result in higher SE improvement.

The cellular network is assumed to be fully loaded consisting of two layers: the cellular layer and the D2D layer. 
There exists $N$ CUEs denoted by $\bm{\mathcal{C}} = \{c_1, c_2, ..., c_N\}$ in the cellular layer {\color{black}where each CUE occupies exactly one CL.} 
There are $M$ D2D pairs in the D2D layer denoted by $\bm{\mathcal{D}} = \{d_1, d_2, ..., d_M\}$ and coexist with the CUEs in the cell and reuse the uplink spectrum in an underlay manner.
Each D2D pair consists of two proximate UEs.
One of the UEs is considered as the D2D transmitter (Tx) while the other one is the D2D receiver (Rx).
{\color{black} It is assumed that the locations of CUEs and D2D pairs are approximated using well known GPS-based \cite{S065} or localization-based \cite{S063,S064} methods.
The transmission of the location data of all CUEs and D2D pairs requires a tiny fraction of the CL capacity considering the typical values of CL bandwidth and user mobility in cellular networks.}
Without loss of generality, half-duplex communication is assumed among the Tx and Rx of a D2D pair since a full-duplex communication consists of two half-duplex communications using two different CLs when the transmitter of the first half-duplex communication is the receiver of the second one and the receiver of the second half-duplex communication is the transmitter of the first one.
It is reasonable to assume that each D2D pair can reuse the CL of exactly one CUE since the Tx module of each UE can transmit at one CL at a time while multiple D2D pairs can share the same CL simultaneously.
The dense cell where the number of D2D pairs is greater than the number of CUEs, i.e., $M>N$, is considered.

\subsection{Interference and Data Rate}

The channel gains of desired links from the Tx of $d_j$ to the Rx of $d_j$ and from $c_i$ to the BS are denoted by $h^d_j$ and $h^c_i$, respectively.
The channel gains of interference links from $c_i$ to the Rx of $d_j$, from the Tx of $d_j$ to the Rx of $d_k$, and from the Tx of $d_j$ to the BS are denoted by $h^{c,d}_{i,j}$, $h^{d,d}_{j,k}$, and $h^{d,b}_j$, respectively.
The power of additive white Gaussian noise is denoted by $\sigma^2$.
The bandwidths of all CLs are assumed to be equal.
The minimum QoS requirements of $c_i$ and $d_j$ in terms of bandwidth normalized data rate (spectral efficiency) are denoted by $R^{c,\rm{min}}_i$ and $R^{d,\rm{min}}_j$, respectively.

There is no inter-layer interference in the cellular layer due to the orthogonality of CLs.
The signal to noise and interference ratios (SINRs) of $c_i$ and $d_j$ are denoted by $\gamma^c_i$ and $\gamma^d_j$, respectively, and expressed as
\begin{align}
\label{CUESINREq}
\gamma^c_i &= \frac{p^c_i h^c_i}{\sigma^2
+\sum\limits_{j=1}^{M} \psi_{i,j} p^d_j h^{d,b}_j}, \\
\label{D2DSINREq}
\gamma^d_j &= \frac{p^d_j h^d_j}{\sigma^2
+\sum\limits_{i=1}^{N} \psi_{i,j} p^c_i h^{c,d}_{i,j} + \sum\limits_{k=1 \atop k\neq j}^{M} \sum\limits_{i=1}^{N}
\psi_{i,j} \psi_{i,k} p^d_k h^{d,d}_{k,j}},
\end{align}
respectively,
where $\psi_{i,j}$ denotes the CUE-D2D resource sharing indicator variable,
$\psi_{i,j}=1$ when $c_i$ and $d_j$ use the same CL; otherwise, $\psi_{i,j}=0$.
The transmission powers of $c_i$ and Tx of $d_j$ are denoted by $p^c_i$ and $p^d_j$, respectively.
The second term in the denominator of \eqref{CUESINREq} corresponds to the intra-layer interference power received at $c_i$ and caused by the D2D pairs that use the CL of $c_i$.
The second and third terms of the denominator of \eqref{D2DSINREq} corresponds to the intra-layer and inter-layer interference powers received at $d_j$, respectively.
The intra-layer and inter-layer interferences are caused by the CUE and D2D pairs that use the same CL of $d_j$, respectively.
The spectral efficiency of $c_i$ and $d_j$ can be formulated as 
\begin{align}
\label{RateC}
R^c_i &= \log_{2}(1+\gamma^c_i),\\
\label{RateD}
R^d_j &=  \log_{2}(1+\gamma^d_j),
\end{align}
respectively.

\subsection{Problem Formulation}
We aim to maximize the overall sum-rate of the cell by joint assignment of CLs and transmission powers to CUEs and underlaying D2D pairs in a distributed manner.
Due to the dense cell assumption, multiple admitted D2D pairs may reuse one CL and some of the D2D pairs may remain unadmitted.
Maximizing the sum-rate of the cell is equivalent to maximizing the overall spectral efficiency of the cell due to the equality of the bandwidths of CLs.
Hence, the sum-rate function which is normalized by the bandwidth can be expressed as
\begin{align}
\label{Eq_Rt}
R^{\rm{Total}} = \sum_{i=1}^N R^c_i + \sum_{j=1}^M \rho_j R^d_j,
\end{align}
where $\rho_j$ is the D2D resource sharing indicator variable,  $\rho_j=1$ when $d_j$ is able to reuse the CL of any CUE;  otherwise, $\rho_j=0$.
The D2D resource sharing indicator can be expressed as
\begin{align}
\label{Eq_Rt2}
&\rho_j = \sum^N_{i=1} \psi_{i,j}, ~~\forall j=1,...,M,\\
\label{PsiDef}
&\psi_{i,j} \in \{0,1\}, ~~\forall i=1,...,N,~~\forall j=1,...,M.
\end{align}

The transmission power vectors of CUEs and D2D pairs can be expressed as 
 $\bm{p}^c = [ p^c_1, p^c_2, ..., p^c_N ]^T $ 
and 
$\bm{p}^d = [ p^d_1, p^d_2, ..., p^d_M ]^T $, respectively.
{\color{black} The resource sharing vector of $d_j$ can be expressed as
$\bm{\psi}_j = [ \psi_{1,j}, \psi_{2,j}, ..., \psi_{N,j}  ]^T$.}
Hence, the  transmission power vector of all CUEs and D2D pairs and the CUE-D2D resource sharing matrix can be expressed as 
$\bm{P} = [ (\bm{p}^c)^T, (\bm{p}^d)^T ]^T$ 
and
$\bm{\Psi} \triangleq [ \bm{\psi}_1, \bm{\psi}_2, ..., \bm{\psi}_M  ]$, respectively.
By expressing the minimum QoS requirements of $c_i$ and $d_j$ in terms of their SINRs as 
\begin{align}
\gamma^{c,\rm{min}}_i &= 2^{ R^{c,\rm{min}}_i } -1, \\
\gamma^{d,\rm{min}}_j &= 2^{ R^{d,\rm{min}}_j } -1,
\end{align}
respectively.

Hence, the resource allocation problem can be formulated as the following mixed-integer non-linear non-convex optimization problem:
\begin{subequations}
\label{OptProbEqInit}
\begin{alignat}{4}
\nonumber
&\max_{\bm{P},\bm{\Psi}} ~~&&R^{\rm{Total}},\\
\nonumber
& \rm{s.t.} &&\eqref{PsiDef},\\
\label{QosCUECon}
& && \gamma_i^c \geq \gamma_i^{c,\rm{min}}, ~~&&&\forall i=1,...,N,\\
\label{QosD2DCon}
& && \gamma_j^d \geq \rho_j \gamma_j^{d,\rm{min}}, ~~&&&\forall j=1,...,M,\\
\label{MaxPcCon}
& && 0 \leq p^c_i \leq P^{c,\rm{max}}_i, ~~&&&\forall i=1,...,N,\\
\label{MaxPdCon}
& && 0 \leq p^d_j \leq P^{d,\rm{max}}_j, ~~&&&\forall j=1,...,M,\\
\label{MaxD2DPerCUE}
& &&\rho_j \in \{ 0, 1 \}, ~~&&&\forall j=1,...,M,
\end{alignat}
\end{subequations}
where the optimization variables are $\bm{P}$ and $\bm{\Psi}$.
The maximum transmission powers of $d_j$ and $c_i$ are denoted by $P^{d,\rm{max}}_j$ and $P^{c,\rm{max}}_i$, respectively.
The minimum QoS constraints of CUEs and D2D pairs are expressed as \eqref{QosCUECon} and \eqref{QosD2DCon}, respectively.
Using the D2D resource sharing indicator variable, D2D admission control is incorporated in constraint \eqref{QosD2DCon}.
The transmission power limits of CUEs and D2D pairs are expressed as \eqref{MaxPcCon} and \eqref{MaxPdCon}, respectively.
Constraint \eqref{MaxD2DPerCUE} guarantees that each D2D pair can share the CL of at most one CUE.
Considering the definition of the D2D resource sharing indicator variable \eqref{Eq_Rt2}, the constraint \eqref{MaxD2DPerCUE} {\color{black} implies that $\bm{\Psi} \in \{0,1\}^{(N \times M)}$ has at most one "1" element in each column which means that each D2D pair can reuse at most one CL.
Hence, there may be several "1" elements in each row of $\bm{\Psi}$ which means that each CL can be shared with multiple D2D pairs.}

{\color{black}
It is important to notice that finding the solution to the optimization problem \eqref{OptProbEqInit} using optimization methods at the BS is considered as  a centralized resource allocation approach} which has a high computational complexity and is not the aim of this research paper.
We aim to solve the the optimization problem \eqref{OptProbEqInit} using a pricing-based game theoretic distributed manner with low computational complexity which is practical for dense D2D deployments.
In other words, we propose a distributed resource allocation algorithm which decomposes the centralized resource allocation algorithm to different phases such that the sum-rate is maximized in a distributed manner while the minimum QoS requirements and the transmission power limit constraints of CUEs and D2D pairs are guaranteed.
The simulation results verify the effectiveness the proposed resource allocation algorithm which is described in the next section.

\section{Distributed Pricing-based Resource Allocation}
\label{PopAlgSec}
In this section, we propose our pricing based game-theoretical algorithm to jointly allocate CLs and transmission powers to CUEs and D2D pairs in a distributed manner in order to maximize the overall sum-rate of the cell.
In order to mitigate the inter-layer interference among CUEs and D2D pairs and the intra-layer interference among D2D pairs, we propose to define a set of price values in order to be used to formulate the interference cost of interfering links which are maintained at the BS.
The game is modeled as a two-phase non-cooperative pricing based game.
CUEs and D2D pairs adjust their transmission powers and CLs in order to maximize their own utility functions in the first phase as explained in Section \ref{SecGame}.
In the second phase, the price values of the interfering links are updated considering the minimum QoS requirements of CUEs and D2D as explained in Section \ref{SecPrice}.
By repeating the two phases as explained in Section \ref{RemarkSec} the sum-rate is maximized as a result of the first phase and the minimum QoS requirements of CUEs and D2D pairs are guaranteed as a result of the second phase.
Hence, the sum-rate is maximized constrained to the minimum QoS requirement of CUEs and D2D pairs in a distributed manner using our two phases proposed algorithm.

\subsection{Non-cooperative Game among CUEs and D2D Pairs}
\label{SecGame}

In order to propose the pricing based resource allocation algorithm, the parameters that represent the price values for the interference links should be introduced.
The price charged for the interference from 
from $c_i$ to the Rx of $d_j$, from the Tx of $d_i$ to the BS, and from the Tx of $d_i$ to the Rx of $d_j$ are denoted by $\theta^{c,d}_{i,j}$, $\theta^{d}_{i}$, and $\theta^{d,d}_{i,j}$, respectively.
The D2D to BS price vector is defined as $\bm{\bar{\theta}}^{d} (i) \triangleq \theta^{d}_{i}$.
The cellular to D2D and D2D to D2D price matrix elements are defined as
$\bm{\theta}^{c,d} (i,j) \triangleq \theta^{c,d}_{i,j}$ and
$\bm{\theta}^{d,d} (i,j) \triangleq \theta^{d,d}_{i,j}$, respectively
and can be expressed in a vector form as
$\bm{\bar{\theta}}^{c,d} \triangleq \rm{vec}( \bm{\theta}^{c,d} )$ and
$\bm{\bar{\theta}}^{d,d} \triangleq \rm{vec}( \bm{\theta}^{d,d})$, respectively
, where $\rm{vec}(.)$ function transforms a matrix into a vector in a column-wise manner.
Hence, the total price vector can be defined as
$\bm{\theta} \triangleq [ (\bm{\bar{\theta}}^{c,d})^T, (\bm{\bar{\theta}}^{d})^T, (\bm{\bar{\theta}}^{d,d})^T ]^T$.

Using the total price vector, the first phase of the proposed algorithm which is the game among CUEs and D2D pairs that can be defined as
\begin{equation}
\label{GameDef}
\mathcal{G}=\{ \bm{\mathcal{K}},  
(\bm{\mathcal{S}}_k)_{k \in \mathcal{K}}, 
(U_k(\bm{\theta},\bm{P},\bm{\Psi})_{k \in \mathcal{K}} )\},
\end{equation}
where $\bm{\mathcal{K}}=\{ \bm{\mathcal{C}}, \bm{\mathcal{D}} \}$ is the set of all CUEs and D2D pairs that are the players of the game.
The strategy of UE $k$ is denoted as $\bm{\mathcal{S}}_k$ which can be formulated as
\begin{equation}
\bm{\mathcal{S}}_k =
\left\{
	\begin{array}{ll}
		\bm{\mathcal{S}}_i^c \triangleq \{ p^c_i, l_i \}, & \mbox{if } k=c_i \in \mathcal{C} \\
		\bm{\mathcal{S}}_j^d \triangleq \{ p^d_j, l_j \}, & \mbox{if } k=d_j \in \mathcal{D} 
	\end{array}
\right. ,
\end{equation}
where $\bm{\mathcal{S}}_i^c$ and $\bm{\mathcal{S}}_j^d$ are the strategies of $c_i$ and $d_j$, respectively.
The transmission powers of $c_i$ and the Tx of $d_j$ are denoted as $p^c_i$ and $p^d_j$, respectively.
The parameters $l_i \in \{1, 2, ..., N\}$ and $l_j \in \{1, 2, ..., N+1 \}$ are the index of the CLs that $c_i$ and $d_j$ use for data transmission, respectively.
Due to the fully loaded assumption of the cell, each $c_i$ uses its own CL and hence, $l_i=i$.
Each D2D pair reuses one CL or remains unadmitted.
Hence, $l_j \in \{1, 2, ..., N+1\}$, where $l_j=k, k \in \{1, 2, ..., N\}$ corresponds to the case where $d_j$ reuses the the CL of $c_k$ and $l_j=N+1$ corresponds to the case where $d_j$ remains unadmitted.
The utility function of the UE $k$ is denoted by $U_k(\bm{\theta},\bm{P},\bm{\Psi})$ and can be expressed as
\begin{equation}
U_k(\bm{\theta},\bm{P},\bm{\Psi}) =
\left\{
	\begin{array}{ll}
		U_i^c(\bm{\theta},\bm{P},\bm{\Psi}), & \mbox{if } k=c_i \in \mathcal{C} \\
		U_j^d(\bm{\theta},\bm{P},\bm{\Psi}), & \mbox{if } k=d_j \in \mathcal{D} 
	\end{array}
\right. ,
\end{equation}
where $U_i^c(\bm{\theta},\bm{P},\bm{\Psi})$ and $U_j^d(\bm{\theta},\bm{P},\bm{\Psi})$ are the utility functions of $c_i$ and $d_j$, respectively.

{\color{black}
By the innovative introduction of $(N+1)^{th}$ CL, the assignment problem can be formulated effectively which will be used to define the D2D index sharing set and to derive the effective closed-form solution to the problem in the next section.}
It should be noted that the following conclusion can also be driven:
\begin{equation}
l_j=N+1 ~\Leftrightarrow ~ \psi_{i,j}=0, \forall i=1, 2, ..., N ~ \Leftrightarrow  ~ \rho_j=0.
\end{equation}

Each UE (CUE or D2D pair) maximizes its payoff function constrained to transmission power limits.
The proposed utility functions of CUEs and D2D pairs and their maximization is discussed in the next sections.

\subsubsection{CUEs}
Our proposed utility function of $c_i$ can be expressed as
\begin{equation}
\label{UtilityFun}
U^c_i(\bm{\theta},\bm{P},\bm{\Psi})=R^c_i(\bm{P},\bm{\Psi}) - p^c_i \sum_{j=1}^M \psi_{i,j}
\theta_{i,j}^{c,d}
\Big( \frac{s(c_i,d_j)}{s(d_j)} \Big)^{-\alpha}
\end{equation}
where the distance between $c_i$ and the Rx of $d_j$ and the distance between the Tx of $d_j$ and its corresponding Rx are denoted by $s(c_i,d_j)$ and $s(d_j)$, respectively.
The first and second terms of \eqref{UtilityFun} are the rate reward and the interference cost of $c_i$, respectively.
It is important to notice that the interference cost is a function of the distances between $c_i$ and other D2D pairs which cause the proposed method to have low signaling overhead when the locations of CUEs and D2D pairs are available.
{\color{black} It should be mentioned that channel-gain-based utility function definition would result in high signaling overhead as there are $N + M + NM + \binom M 2$ channel gain values in the network, while the distance-based utility function definition which requires $N+M$ location values is much more suitable from the signaling overhead point of view and can be considered as a practical approach.}
Each CUE aims to maximize its payoff function by adjusting its transmission power which can be formulated as
\begin{subequations}
\label{OptProbEq}
\begin{alignat}{4}
&{p^c_i}^* = &&\argmax_{p^c_i}~~ &&&& U^c_i(\bm{\theta},\bm{P},\bm{\Psi}),\\
\label{Const1}
& &&~~~~~\rm{s.t.}  &&&& 0 \leq p^c_i \leq P^{c,\rm{max}}_i,
\end{alignat}
\end{subequations}
where ${p^c_i}^*$ is the optimal transmission power strategy of $c_i$.
The optimization variable of \eqref{OptProbEq} is $p^c_i$ as $c_i$ can only adjust its transmission power and is not able to change the transmission powers of other CUEs and D2D pairs, the resource sharing indicator variables of D2D pairs, and the interference cost which is set by the BS.
Therefore, the utility function $U^c_i(\bm{\theta},\bm{P},\bm{\Psi})$ can be expressed as a function of $p^c_i$ as
$U^c_i(\bm{\theta},\bm{P},\bm{\Psi}) \triangleq f^c_i( p^c_i)$.

We define the D2D sharing index set $\bm{\beta}_i, (i=1, 2, ..., N)$ as the set of indices of the D2D pairs that reuse the CL of $c_i$ which is formulated as
$\bm{\beta}_i \triangleq \{ \beta_{i,1}, \beta_{i,2}, ..., \beta_{i,N_i} \}$,
where $N_i$ and $\beta_{i,k}$ are the number of D2D pairs and the index of $k^{th}$ D2D pair that reuse the CL of $c_i$, respectively.
Since some D2D pairs may remain unadmitted, we define the set of unadmitted D2D pairs as D2D sharing index set
$\bm{\beta}_{N+1} \triangleq \{ \beta_{i,1}, \beta_{i,2}, ..., \beta_{i,N_{N+1}} \}$,
where $N_{N+1}$ is the number of unadmitted D2D pairs.
Considering the definition of $\bm{\beta}_i$ it can be concluded that
\begin{subequations}
\begin{align}
& \bigcup_{i=1}^{N+1} \bm{\beta}_i = \mathcal{D}, \\
& \bm{\beta}_i \cap \bm{\beta}_j = \emptyset, ~ \forall i \in \{1, 2, ..., N+1\} \neq j \in \{1, 2, ..., N+1\}.
\end{align}
\end{subequations}

Since the D2D sharing index set $\bm{\beta}_i$ consists of all the D2D pairs that reuse the CL of $c_i$, the utility function of \eqref{UtilityFun} can be simplified by removing the D2D and CUE-D2D resource sharing indicator variables as
\begin{equation}
\label{UtilC}
f^c_i( p^c_i) = \log_{2} (1+p^c_i \Gamma^c_i) - p^c_i q^c_i,
\end{equation}
where $\Gamma^c_i$ and $q^c_i$ are constant with respect to $p^c_i$ and can be formulated as
\begin{align}
\label{GammaC}
\Gamma^c_i &= 
\frac{h^c_i}{\sigma^2+\sum\limits_{j=1}^{N_i} p^d_{\beta_{i,j}} h^{d,b}_j},\\
\label{qC}
q^c_i &= \sum_{j=1}^{N_i} \theta_{i,\beta_{i,j}}^{c,d} 
\Big( \frac{s(c_i,d_{\beta_{i,j}})}{s(d_{\beta_{i,j}})} \Big)^{-\alpha},
\end{align}
respectively.
The function $f^c_i( p^c_i)$ is strictly concave with respect to $p^c_i$ since the first and second terms of \eqref{UtilC} are strictly concave and linear with respect to $p^c_i$, respectively. 
Due to the linearity of \eqref{Const1} with respect to $p^c_i$, the feasible area of \eqref{OptProbEq} is convex.
Considering the concavity of \eqref{UtilC}, the optimization problem \eqref{OptProbEq} is also convex and can be solved analytically. 
The closed-form solution can be formulated as
\begin{equation}
\label{OptCPow}
{p^c_i}^*= \mathcal{F}(q^c_i,\Gamma^c_i,P^{c,\rm{max}}_i),
\end{equation}
where
\begin{equation}
\mathcal{F}(x, y, z) \triangleq
\left\{
	\begin{array}{ll}
		0, 
		&\mbox{if}~ (\frac{1}{\ln(2)x} - \frac{1}{y}) \leq 0 \\
		\frac{1}{\ln(2)x} - \frac{1}{y}, ~
		&\mbox{if}~ 0 < (\frac{1}{\ln(2)x} - \frac{1}{y}) < z \\
		z, 
		&\mbox{if}~ (\frac{1}{\ln(2)x} - \frac{1}{y}) \leq 0
	\end{array}
\right.,
\end{equation}
and the proof is {\color{black} presented in appendix} \ref{AppA}.

\subsubsection{D2D Pairs}
If $d_j$ reuses the CL of $c_i$, our proposed utility function for $d_j$ can be formulated as
\begin{equation}
\label{UtilityFunD}
U^d_j(\bm{\theta},\bm{P},\bm{\Psi})=\rho_j R^d_j(\bm{P},\bm{\Psi}) - p^d_j \Big(
\sum_{i=1}^{N} \psi_{i,j} \theta^{d}_{j} 
\Big(\frac{s(d_j,BS)}{s(c_i)} \Big)^{-\alpha}
+ 
\sum_{i=1}^{N} \sum\limits_{k=1 \atop k\neq j}^M \psi_{i,j} \psi_{i,k} 
\theta_{j,k}^{d,d} 
\Big(\frac{s(d_j,d_k)}{s(d_k)}\Big)^{-\alpha}
 \Big),
\end{equation}
where the distance between the Tx of $d_j$ and the BS, the distance between $c_i$ and its corresponding receiver which is the BS, and the distance between the Tx of $d_j$ and the Rx of $d_k$ are denoted by 
$s(d_j,BS)$, $s(c_i)$, and $s(d_j,d_k)$, respectively.
The first and second terms of \eqref{UtilityFunD} are the rate reward and the interference cost of $d_j$, respectively.
The proposed cost term is a function of the locations of D2D pairs and CUEs.
It should be mentioned that the rate reward depends on the admission of $d_j$ and hence is multiplied by $\rho_j$.
Each D2D pair aims to maximize its utility function by adjusting its transmission power and CL which can be formulated as the following mixed-integer programming:
\begin{subequations}
\label{OptProbEqD}
\begin{alignat}{4}
&{\{  {p^d_j}^*, {\bm{\psi}_j}^*  \}} = && \argmax_{p^d_j, \bm{\psi}_j} ~~&&&& U^d_j(\bm{\theta},\bm{P},\bm{\Psi}),\\
\label{Const1D}
& &&~~~~~\rm{s.t.}  &&&& 0 \leq p^d_j \leq P^{d,\rm{max}}_j, \\
& && &&&& \sum^N_{i=1} \psi_{i,j} \leq 1,\\
& && &&&& \psi_{i,j} \in \{0, 1\},
\end{alignat}
\end{subequations}
where ${p^d_j}^*$ and ${\bm{\psi}_j}^*$ are the optimal transmission power and resource sharing indicator variable of $d_j$, respectively, that correspond to the optimal strategy of $d_j$.
Since the optimization problem \eqref{OptProbEqD} is a mixed-integer problem, each D2D pair may update its strategy according to one of the following three CL updating schemes:
\begin{itemize}
\item Keeping CL: 
The D2D pair is using a specific CL for data transmission.
By optimizing the utility function of the D2D pair and considering all the other CLs, it is concluded that reusing the other CLs does not result in a higher payoff compared to reusing the current one.
Hence, the D2D pair keeps its current CL.
\item Changing CL:
The highest payoff of the D2D pair can be obtained by reusing a CL other than the current one that the D2D pair is reusing.
Hence, the CL of the D2D pairs is changed to the new CL.
\item Un-admission: 
An unadmitted D2D pair is not able to reuse any of the CLs and its transmission power should remain zero to avoid interference to other CUEs and D2D pairs, and hence its data rate is equal to zero.
Considering \eqref{UtilityFunD}, it can be concluded that the payoff of an unadmitted D2D pair is equal to zero.
\end{itemize}

In order to solve the mixed-integer problem \eqref{OptProbEqD}, the D2D sharing index sets and CL strategy updating schemes can be used to simplify the D2D utility functions.
Thus, the utility function of $d_j$ which is expressed as \eqref{UtilityFunD} can be simplified as $f^d_j(p^d_j)$ and $\bar{f}^d_{j,i} (p^d_j)$ when the CL strategy updating scheme of $d_j$ is Keeping CL and Changing CL, respectively.
The utility function of $d_j$ when the D2D pair changes its CL to the CL of $c_i$ is denoted by $\bar{f}^d_{j,i} (p^d_j)$.
As the simplified utility functions $f^d_j(p^d_j)$ and $\bar{f}^d_{j,i} (p^d_j)$ are formulated in the following, it can be seen that the resource sharing indicator variables are removed from the initial utility function definition \eqref{UtilityFunD} as a result of using D2D sharing index sets.
The simplified utility function of $d_j$ is a function of the D2D transmission power $p^d_j$ as a result of considering the cL strategy updating schemes.

The utility function of $d_j$ when it is using the CL of $c_i$ and its CL strategy updating scheme is Keeping CL can be formulated as
\begin{equation}
\label{UtilityDKeep}
f^d_j(p^d_j) = \log_{2} (1+p^d_j \Gamma^d_j) - p^d_j q^d_j,
\end{equation}
where $j \in \bm{\beta}_i$.
The parameters $\Gamma^d_j$ and $q^d_j$ are constant with respect to $p^d_j$ and can be expressed as
\begin{align}
\label{GammaDKeep}
\Gamma^d_j &=
 \frac{h^d_j}{\sigma^2+p^c_i h^{c,d}_{i,j} + \sum\limits_{k=1 \atop \beta{i,k}\neq j}^{N_i} p^d_{\beta_{i,k} } h^{d,d}_{\beta_{i,k},j}}, \\
\label{qDKeep}
q^d_j &=
\theta^{d}_{j} 
\Big(\frac{s(d_j,bs)}{s(c_i)} \Big)^{-\alpha}
+ \sum\limits_{k=1 \atop \beta_{i,k}\neq j}^{N_i}
\theta^{d,d}_{j,\beta_{i,k}} 
\Big(\frac{s(d_j,d_{\beta_{i,k}})}{s(d_{\beta_{i,k}})}\Big)^{-\alpha},
\end{align}
respectively.
Using a similar approach to the previous section, the optimal power of $d_j$ when its CL is not changed can be expressed using a closed-form solution as
\begin{equation}
\label{OptDPowKeep}
{p^d_j}^*= \mathcal{F}(q^d_j,\Gamma^d_j,P^{d,\rm{max}}_j),
\end{equation}
where the proof is similar to the proof of \eqref{OptCPow}.

Similarly, the utility function of $d_j$ when its CL is changed to the CL of $c_i$ can be expressed as
\begin{equation}
\label{UtilityDChange}
\bar{f}^d_{j,i} (p^d_j) = \log_2(1+p^d_j\bar{\Gamma}^d_{j,i}) - p^d_j\bar{q}^d_{j,i},
\end{equation}
where $j \notin \bm{\beta}_i$.
The parameters $\bar{\Gamma}^d_{j,i}$ and $\bar{q}^d_{j,i}$ are constant with respect to $p^d_j$ and can be expressed as
\begin{align}
\label{GammaDChange}
\bar{\Gamma}^d_{j,i} &=
 \frac{h^d_j}{\sigma^2+p^c_i h^{c,d}_{i,j} + \sum\limits_{k=1}^{N_i} p^d_{\beta_{i,k} } h^{d,d}_{\beta_{i,k},j}}, \\
\label{qDChange}
\bar{q}^d_{j,i} &=
\theta^{d}_{j} \Big(\frac{s(d_j,bs)}{s(c_i)} \Big)^{-\alpha}
 + \sum\limits_{k=1}^{N_i}
\theta^{d,d}_{j,\beta_{i,k}} \Big(\frac{s(d_j,d_{\beta_{i,k}})}{s(d_{\beta_{i,k}})}\Big)^{-\alpha},
\end{align}
respectively.
Hence, the optimal power of $d_j$ when the its CL is changed to the CL of $c_i$ can be formulated as
\begin{equation}
\label{OptDPowChange}
{p^d_j}^*=\mathcal{F}(\bar{q}^d_{j,i}, \bar{\Gamma}^d_{j,i}, P^{d,\rm{max}}_j),
\end{equation}
which is a closed-form solution and its proof is similar to the proof of \eqref{OptCPow}.

Each D2D pair computes its payoff value considering the three CL updating schemes.
The transmission powers of the Keeping CL and Changing CL schemes have been derived in a closed-form, and hence their payoff values can be computed from the D2D pair utility function expression \eqref{UtilityFunD}.
One and $(N-1)$ payoff values are computed for the Keeping CL and Changing CL schemes, respectively.
{\color{black}From the D2D utility function definition it can be easily verified that the payoff value of the Un-admission scheme is equal to zero.
Most of the admission strategy evaluation metrics in the literature such as the proposed admission strategy of \cite{R003} are based on complex formulas or formulas that are different from the proposed resource allocation method and result in additional computational complexity of the algorithm.
Our proposed admission strategy is to compare the payoff value with zero which utilizes the novel D2D utility function definition and hence is a simple, fast, and novel admission strategy.}
In other words, the payoff values of both Keeping CL and Changing CL schemes are computed and compared to zero.
The CL corresponding to the maximum payoff value is selected as the optimal CL.
If all the payoff values corresponding to both Keeping CL and Changing CL schemes of a D2D pair are less than zero, the D2D pair will select the Un-admission scheme with zero payoff value which is our proposed D2D admission strategy suitable for dense D2D deployment in beyond 5G networks.

\subsection{Pricing Considering QoS Constraints}
\label{SecPrice}

The interference cost should be updated by adjusting the price vector $\bm{\theta}$.
Each CUE and D2D pair request a minimum QoS in terms of data rate from the network.
If the achievable data rate of a CUE or a D2D pair is lower than its minimum required rate, we propose to increase those elements of the price vector that are corresponding to the interference links of that CUE or D2D pair.
The price of the interference links can be decreased if the data rate is much larger than the minimum required rate.
The pricing vector update procedure depends on the QoS of CUEs and D2D pairs.

\subsubsection{QoS of CUEs}
After transmission power updating of CUEs and D2D pairs that is described in Section \ref{SecGame}, the QoS constraints of CUEs should be investigated to adjust the price  vector $\bm{\theta}$.
All $N_i$ D2D pairs with indices $\bm{\beta}_i$ reuse the CL of $c_i$ and cause interference to it.
The data rate of $c_i$ in $n^{th}$ iteration denoted as ${R^c_i}^{(n)}$, should be compared to the minimum QoS requirement of $c_i$, i.e., $R^{c,\rm{min}}_i$, at the BS.
If ${R^c_i}^{(n)} < R^{c,\rm{min}}_i$, we propose to increase the price of the links that cause interference to $c_i$ in the $(n+1)^{th}$ iteration as
\begin{equation}
\label{PriceUpEqC}
{\theta^{d}_{\beta_{i,j}}}^{(n+1)} = {\theta^{d}_{\beta_{i,j}}}^{(n)} (1+\lambda_1), ~j=1,...,N_i,
\end{equation}
where ${\theta^{d}_{\beta_{i,j}}}^{(n+1)}$ and ${\theta^{d}_{\beta_{i,j}}}^{(n)}$ are the price of the interference links from $d_{\beta_{i,j}}$ to the BS in the $(n+1)^{th}$ and $n^{th}$ iterations, respectively.
The price updating parameter for the price increment scheme is denoted by $\lambda_1$.

Additionally, we propose to decrease the price of interference links, if the data rate is larger than minimum required data rate plus a marginal rate.
In other words, if ${R^c_i}^{(n)}  > R^{c,\rm{min}}_i (1+\zeta)$, we propose to update the price of interference links as
\begin{equation}
\label{PriceDownEqC}
{\theta^{d}_{\beta_{i,j}}}^{(n+1)} = {\theta^{d}_{\beta_{i,j}}}^{(n)} (1-\lambda_2), ~j=1,...,N_i,
\end{equation}
where $\zeta$ is the marginal rate adjustment parameter and $\lambda_2$ is the price updating parameter for the price decrement scheme.

\subsubsection{QoS of D2D Pairs}
Similar to the pricing procedure of CUEs, the data rate of $d_j$ in $n^{th}$ iteration denoted as ${R^d_j}^{(n)}$, should be compared to the minimum required data rate of $d_j$, i.e., $R^{d,\rm{min}}_j$, at its receiver.
Assuming $d_j$ to be using the CL of $c_i$, $(N_i-1)$ D2D pairs with indices 
$\bar{\bm{\beta}}_i = \bm{\beta}_i - \{k | \beta_{i,k}=j\}$
and $c_i$ cause interference to $d_j$.
If ${R^d_j}^{(n)} < R^{d,\rm{min}}_j$, we propose to update the price of the interference links as
\begin{align}
\label{PriceUpEqD1}
{\theta^{c,d}_{i,j}}^{(n+1)} &= {\theta^{c,d}_{i,j}}^{(n)} (1+\lambda_1),\\
\label{PriceUpEqD2}
{\theta^{d,d}_{k,j}}^{(n+1)} &= {\theta^{d,d}_{k,j}}^{(n)} (1+\lambda_1), ~k \in \bar{\bm{\beta}}_i .
\end{align}

Additionally, if ${R^d_j}^{(n)} > R^{d,\rm{min}}_j (1+\zeta)$, we proposed price updating of the interference links can be expressed as
\begin{align}
\label{PriceDownEqD1}
{\theta^{c,d}_{i,j}}^{(n+1)} &= {\theta^{c,d}_{i,j}}^{(n)} (1-\lambda_2),\\
\label{PriceDownEqD2}
{\theta^{d,d}_{k,j}}^{(n+1)} &= {\theta^{d,d}_{k,j}}^{(n)} (1-\lambda_2), k\in \bar{\bm{\beta}}_i .
\end{align}

\section{Proposed Algorithm}
\label{RemarkSec}
In this section, we propose our resource allocation algorithm named Distributed Spectrally Efficient Resource Allocation (DSERA) algorithm from an algorithmic point of view.
In addition to that, some remarks related to the DSERA algorithm are also discussed.

\subsection{Two Phases of the Resource Allocation Algorithm}
Our proposed resource allocation algorithm consists of two phases that are already explained in the previous sections.
It should be noted that the first phase of the proposed algorithm results in a distributed overall sum-rate maximization since the data rate is considered as the reward term and the interference is modeled as the cost term of the utility function.
The maximum and minimum transmission power constraints are considered in the optimization process of the first phase.
The price values of the interfering links are adjusted in the second phase according to the QoS constraints of CUEs and D2D pairs.
The repeated competition among CUEs and D2D pairs to reuse the uplink spectrum considering the set of prices that are maintained by the BS will eventually result in a resource allocation that guarantees the minimum QoS requirements of CUEs and admitted D2D pairs.

\subsection{Signaling Overhead}
The signaling overhead is an important issue in wireless systems.
In our proposed method, the channel gains of the links from the CUEs to the BS and from the Txs of D2D pairs to their corresponding Rxs which are the desired links should be estimated.
In order to compute the data rate of $c_i$, the BS measures the aggregate interference of the CL corresponding to $c_i$ and computes the data rate of $c_i$ using $h^c_i$ and $p^c_i$.
Therefore, the payoff value of $c_i$ is also computed.
In order to compute the data of $d_j$, the Rx of $d_j$ measures the aggregate interference of the CL corresponding to $d_j$ and computes its data rate and payoff value using $p^d_j$ and $h^d_j$.
The channel gains of other links are not required to compute the data rate as the aggregate interference is computed at each receiver and the proposed utility function utilizes a distance-based cost term.
Hence, the final proposed algorithm has a low signaling overhead.
Therefore, our proposed algorithm is a practical resource allocation scheme for dense scenarios.

\subsection{Price Updating Procedure}
We propose two price updating schemes which are whole price updating and step-by-step price updating schemes and are introduced in the following:
\begin{itemize}
\item
\textbf{Whole price updating}: In this scheme the price values of all the links are updated simultaneously at the evaluation of phase two of the proposed algorithm, and hence is quicker compared to the other price updating strategy.
\item
\textbf{Step-by-step price updating}: In this scheme the price values corresponding to one CUE and the D2D pairs that are using the CL of the CUE are updated at each evaluation of the proposed algorithm which is slower compared to the other price updating strategy.
\end{itemize}
Due to the distributed manner of our proposed resource allocation algorithm, the step-by-step price updating scheme which updates a part of the prices at each evaluation of phase two of the proposed algorithm is expected to results in a higher sum-rate compared to the sum-rate of the whole price updating scheme.
The effect of the price updating procedure is investigated using numerical evaluation.

\subsection{Algorithmic Representation of the First Phase}
The first phase of the DSERA algorithm which is a non-cooperative game among CUEs and D2D pairs has been discussed in Section \ref{SecGame} and summarized as Algorithm \ref{TableFirstPhase}.

\begin{algorithm}[!h]
\caption{First Phase of the DSERA Algorithm}
\label{TableFirstPhase}
\begin{algorithmic}[1]
\For {$i \in \mathcal{C}$}
\State Compute $\Gamma^c_i$ from \eqref{GammaC}
\State Compute $q^c_i$ from \eqref{qC}
\State Compute $p^c_i={p^c_i}^*$ from \eqref{OptCPow}
\EndFor
\For {$j \in \mathcal{D}$}
\State $\tilde{u}^d_j = \{0, ..., 0\}^T \in \{0, 1\}^{(N+1) \times 1}$
\State $\tilde{p}^d_j = \{0, ..., 0\}^T \in \mathcal{R}^{(N+1) \times 1}$
	\For {$i \in \mathcal{C}$}
		\If {$\psi_{i,j}=1$}
\State Compute $\Gamma^d_j$ from \eqref{GammaDKeep}
\State Compute $q^d_j$ from \eqref{qDKeep}
\State Compute $\tilde{p}^d_j(i) = {p^d_j}^*$ from \eqref{OptDPowKeep}
\State Compute $\tilde{u}^d_j(i) = f^d_j(p^d_j)$ from \eqref{UtilityDKeep}

\Else
\State Compute $\bar{\Gamma}^d_{j,i}$ from \eqref{GammaDChange}
\State Compute $\bar{q}^d_{j,i}$ from \eqref{qDChange}
\State Compute $\tilde{p}^d_j(i) = {p^d_j}^*$ from \eqref{OptDPowChange}
\State Compute $\tilde{u}^d_j(i) = \bar{f}^d_{j,i} (p^d_j)$ from \eqref{UtilityDChange}
		\EndIf
	\EndFor
	\State $l_j = \arg \max_{k \in \{0, 1\}^{(N+1)}} \tilde{u}^d_j$
	\State $p^d_i = \tilde{p}^d_j(l_j)$
\EndFor
\end{algorithmic}
\end{algorithm}
\begin{algorithm}[!h]
\caption{Second Phase of the DSERA Algorithm - Whole price updating scheme}
\label{TableSecondPhaseWhole}
\begin{algorithmic}[1]
\State Given $n$
\For {$i \in \mathcal{C}$}
\State Compute ${R^c_i}^{(n)}=R^c_i$ from \eqref{RateC}
\If {${R^c_i}^{(n)} < R^{c,\rm{min}}_i$}
\State Update price vector from \eqref{PriceUpEqC}
\ElsIf {${R^c_i}^{(n)}  > R^{c,\rm{min}}_i (1+\zeta)$}
\State Update price vector from \eqref{PriceDownEqC}
\EndIf
\EndFor
\For {$j \in \mathcal{D}$}
\State Compute ${R^d_j}^{(n)}=R^d_j$ from \eqref{RateD}
\If {${R^d_j}^{(n)} < R^{d,\rm{min}}_j$}
\State Update price vector from \eqref{PriceUpEqD1} and \eqref{PriceUpEqD2}
\ElsIf {${R^c_i}^{(n)}  > R^{c,\rm{min}}_i (1+\zeta)$}
\State Update price vector from \eqref{PriceDownEqD1} and \eqref{PriceDownEqD2}
\EndIf
\EndFor
\end{algorithmic}
\end{algorithm}

\begin{algorithm}[!h]
\caption{Second Phase of the DSERA Algorithm - Step-by-step price updating scheme}
\label{TableSecondPhaseStep}
\begin{algorithmic}[1]
\State Given $n$ and $i$
\State Compute ${R^c_i}^{(n)}=R^c_i$ from \eqref{RateC}
\If {${R^c_i}^{(n)} < R^{c,\rm{min}}_i$}
\State Update price vector from \eqref{PriceUpEqC}
\ElsIf {${R^c_i}^{(n)}  > R^{c,\rm{min}}_i (1+\zeta)$}
\State Update price vector from \eqref{PriceDownEqC}
\EndIf
\For {$j \in \bm{\beta}_i$}
\State Compute ${R^d_j}^{(n)}=R^d_j$ from \eqref{RateD}
\If {${R^d_j}^{(n)} < R^{d,\rm{min}}_j$}
\State Update price vector from \eqref{PriceUpEqD1} and \eqref{PriceUpEqD2}
\ElsIf {${R^c_i}^{(n)}  > R^{c,\rm{min}}_i (1+\zeta)$}
\State Update price vector from \eqref{PriceDownEqD1} and \eqref{PriceDownEqD2}
\EndIf
\EndFor
\end{algorithmic}
\end{algorithm}
\begin{algorithm}[!h]
\caption{Distributed Spectrally Efficient Resource Allocation (DSERA) algorithm}
\label{TableDSERA}
\begin{algorithmic}[1]
\State Set counter $n=1$, $i=1$, initialize $\bm{\theta}(0)$ and $\bm{P}(0)$
\While {$\{\exists k\in \mathcal{C} ~\textrm{s.t.}~{R^c_k}^{(n)} < R^{c,\rm{min}}_k \}~\textrm{OR} ~ \{ \exists k \in \mathcal{D} ~\textrm{s.t.} ~{R^d_k}^{(n)} < R^{d,\rm{min}}_k \}$ }
\If {${R^c_i}^{(n)} < R^{c,\rm{min}}_i ~\textrm{AND} ~ {R^d_j}^{(n)} < R^{d,\rm{min}}_j, \forall j\in \bm{\beta}_j$}
\State $i=\textrm{mod}(i,N)+1$ and \textbf{Continue}
\EndIf
\While {$||\bm{P}(n)-\bm{P}(n-1)|| > \epsilon$}
\State First phase of the DSERA algorithm as described in Algorithm \ref{TableFirstPhase}.
\EndWhile
\State Second phase of the DSERA algorithm as described in Algorithm \ref{TableSecondPhaseStep}.
\State $n=n+1$, $i=\textrm{mod}(i,N)+1$
\EndWhile 
\end{algorithmic}
\end{algorithm}

\subsection{Algorithmic Representation of the Second Phase}
Interference price vector updating considering the QoS constraints is the second phase of our proposed resource allocation algorithm.
The algorithmic procedure of this phase of the DSERA algorithm depends on the utilized price updating scheme.
This step considering the whole and step-by-step price updating schemes are described as Algorithms \ref{TableSecondPhaseWhole} and \ref{TableSecondPhaseStep}, respectively.

\subsection{Proposed Resource Allocation Algorithm}
The final resource allocation algorithm is the repetition of the introduced first and second phases.
Our proposed DSERA algorithm can be summarized as Algorithm \ref{TableDSERA}.
It should be noted that our proposed price updating procedure is the step-by-step price updating scheme as it has been used in the DSERA algorithm.
The price values at $n^{th}$ iteration is also denoted by $\bm{P}(n)$.

\subsection{Computational Complexity Analysis}
{\color{black} Due to the time variant channel gains and the mobility of users, the resource allocation algorithm has to be repeated once in a while to find the subchannel allocation and transmission powers of CUEs and D2D pairs.
During each run of the resource allocation algorithm, the channel gains as well as the locations of the users are assumed to be constant.
Our system model consists of three type of computation nodes that are the D2D receivers, the BS acting as the receiver of CUEs, and the BS acting as the central computation node.
We present the computational complexity of each computation node in the following.}
\begin{itemize}
\item
{\color{black}
BS acting as the central computation node:

This computation node updates the price values and computes the constant values.
Considering the locations of CUEs and D2D pairs, the distance values denoted by $s(c_i,bs)$, $s(c_i,d_j)$,$s(d_i,d_j)$, and $s(d_i,bs)$ and the ratio of the distance values to the power of $(-\alpha)$ denoted by
$\Big( \frac{s(c_i,d_{\beta_{i,j}})}{s(d_{\beta_{i,j}})} \Big)^{-\alpha}$,
$\Big(\frac{s(d_j,bs)}{s(c_i)} \Big)^{-\alpha}$,
and 
$\Big(\frac{s(d_j,bs)}{s(c_i)} \Big)^{-\alpha}$
which are used in equations \eqref{qC}, \eqref{qDKeep}, and \eqref{qDChange} are constants.
The constant values of the algorithm can be computed just once to avoid high computational complexity and these values should be computed at the BS acting as the central computation node.
Hence, $2(M+1)(M+N)$ multiplications, $M^2+2MN$ divisions, and $M^2+2MN$ powers are required to compute the ratios of the distances to the power of $(-\alpha)$.
Considering the highest computational complexity operation, the computational complexity of computing the constant values is $O( M(M+N) )$ which is performed at the BS acting as a central computation node.

Considering the equations \eqref{PriceUpEqC} and \eqref{PriceDownEqC} the price updating procedure of each CUE consists of $N_i$ multiplications.
Assuming the upper bound average value of $N_i$ to be $M/N$, the number of required multiplications for the price updating of CUEs would be bounded by $K_1 N (M/N) = K_1M$ where $K_1$ represents the number of iterations of the outer loop of Algorithm \ref{TableDSERA}.
Considering the equations \eqref{PriceUpEqD1}, \eqref{PriceUpEqD2}, \eqref{PriceDownEqD1}, and \eqref{PriceDownEqD2} the price updating procedure of each D2D pair consists of $1+(N_i-1)$ multiplications.
Hence, the number of required multiplications for the price updating of D2D pairs would be bounded by $K_1 M (M/N)= K_1 M^2/N$.
Hence, the computational complexity of the price updating procedure is $O(K_1 M^2/N)$.

It can be concluded that the computational complexity of the BS acting as a central computation node is $O(M(M+N))$ since the price updating procedure is based on multiplications but the computation of constant values is based on divisions.
}
\item
{\color{black}
BS acting as the receiver of CUEs:

The equations 
\eqref{GammaC}, \eqref{qC}, and \eqref{OptCPow} 
should be computed at the BS acting as the receiver of CUEs.
Considering the computation of the constants at the BS acting as the central computation node 
and the computation of the aggregate interference, 
there exist $N_i$ 
and three divisions for each CL at 
each iteration of the game.
Hence, 
the computational complexity of the BS acting as the receiver of CUEs 
would be bounded by 
$O(K_1 K_2 M/N* N) = O( K_1 K_2 M)$ 
where 
$K_2$ 
represents the number of iterations of the inner loop of 
Algorithm \ref{TableDSERA}.
}
\item
{\color{black}
D2D receivers:

The equations \eqref{GammaDKeep}, \eqref{qDKeep}, \eqref{OptDPowKeep}, \eqref{GammaDChange}, \eqref{qDChange}, and \eqref{OptDPowChange} 
should be computed at the D2D receivers.
Considering the computation of the constants at the BS acting as the central computation node and the computation of the aggregate interference, there exist 
$1+ (N_i-1)$ 
multiplications and three divisions for the computation of equations 
\eqref{GammaDKeep}, \eqref{qDKeep}, and \eqref{OptDPowKeep} 
followed by 
$(N-1) (N_i+1)$ 
multiplications and $3N_i$ divisions for the computation of equations 
\eqref{GammaDChange}, \eqref{qDChange}, and \eqref{OptDPowChange}. 
Hence, the computational complexity of D2D receivers would be 
$O(K_1 K_2 M N_i N) = O(K_1 K_2 M^2)$.}
\end{itemize}
{\color{black}
It should be noted that computation of constants can be implemented using very fast methods 
based on the usage of look-up tables on hardware. 
These methods consume more storage 
but reduce the computational complexity dramatically.
As a result, 
It can be concluded that 
the computational complexity of the proposed method is 
$O(K_1 K_2 M^2)$.}

\section{Numerical Results}
\label{SimulationSec}
The performance of our proposed algorithm is evaluated in this section through computer simulations.
We consider a fully loaded single cell network with uniformly distributed CUEs and D2D-Txs.
Each D2D-Rx is uniformly distributed in a cluster around its corresponding transmitter.
The path loss model of \cite{R042_026} is used.
{\color{black} The minimum QoS requirements of the CUEs and D2D pairs are assumed to have a uniform distribution as it has been used in a wide range of research papers such as \cite{R003,R042_026,R146}. }
The main simulation parameters are listed in Table \ref{TableSymParam}.
The network performance evaluation metric is the sum-rate of CUEs and admitted D2D pairs.
We investigate the sum-rate and the average rate of different utility functions to verify the effectiveness of our proposed utility function.
In order to demonstrate the higher SE performance of our proposed scheme, the performance of the proposed algorithm is compared to other existing methods for different numbers of CLs and deployment densities of D2D pairs.

\begin{table}[!h]
\begin{center}
\caption{Simulation Parameters} \label{TableSymParam}
\def\arraystretch{1.5}
\begin{tabular}{|p{0.9in}|p{1in}|p{2in}|} \hline
\multicolumn{2}{|p{2.0in}|}{\textbf{Parameter}}& \textbf{Value}  \\ \hline
\multicolumn{2}{|p{1.5in}|}{Physical link type} & {Uplink} \\ \hline
\multicolumn{2}{|p{1.5in}|}{Cell radius} & {400 m} \\ \hline
\multicolumn{2}{|p{1.5in}|}{Noise power ($\sigma^2_N$)} & -114 dBm \\ \hline
\multicolumn{2}{|p{1.5in}|}{Path loss model} & $15.3+37.6 \log_{10}D$ (\textit{D} in m) \\ \hline
\multicolumn{2}{|p{2in}|}{Max D2D Tx power ($P^{d,\rm{max}}$)} & 21 dBm\\ \hline
\multicolumn{2}{|p{1.5in}|}{Max CUE power ($P^{c,\rm{max}}$)} & 24 dBm\\ \hline
\multicolumn{2}{|p{2.0in}|}{CUE and D2D min QoS
($\gamma^{c,\rm{min}}$)}
& Uniformly distributed in [0, 10] dB\\ \hline
\multicolumn{2}{|p{2in}|}{Shadowing standard deviation} & 8 dB\\ \hline
\multicolumn{2}{|p{1.5in}|}{D2D cluster radius} & Uniformly distributed in [10, 40] m\\ \hline
\multicolumn{2}{|p{1.5in}|}{Number of CUEs (N)} & 6, 8, ..., 16 \\ \hline
\multicolumn{2}{|p{1.5in}|}{Number of D2D pairs (M)} & 4N, 6N, ..., 20N \\ \hline
\end{tabular}
\end{center}
\vspace*{-0.6cm}
\end{table}

\subsection{Utility Function Definition and Price Updating Procedure}

\begin{figure}[!h]
    \centering
    \includegraphics[trim={0cm 0 0cm 0},clip,width=0.6\linewidth]{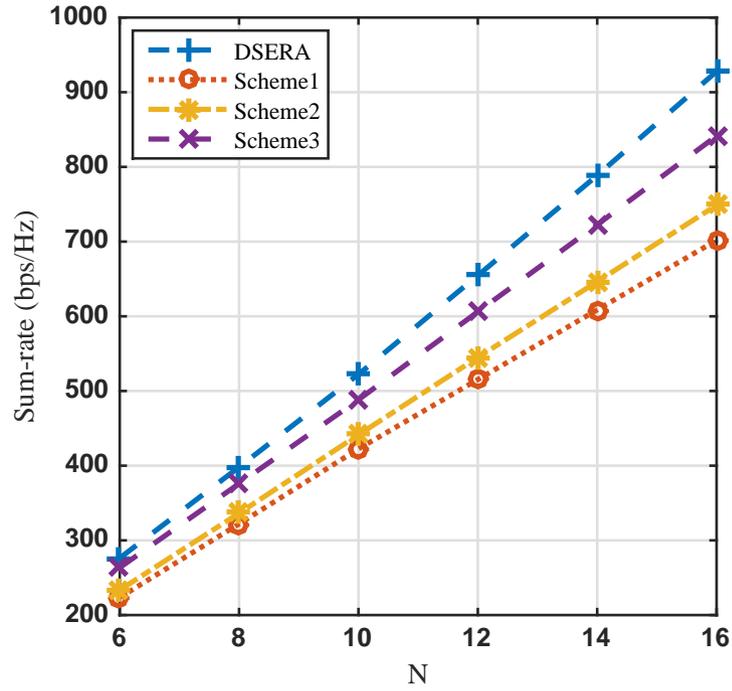}
    \caption{Sum-rate of different utility function definitions and price updating procedures versus the number of CLs when $M/N=20$.}
    \label{FigSumDistSchVsN}
\end{figure}
\begin{figure}[!h]
    \centering
    \includegraphics[trim={0cm 0 0cm 0},clip,width=0.6\linewidth]{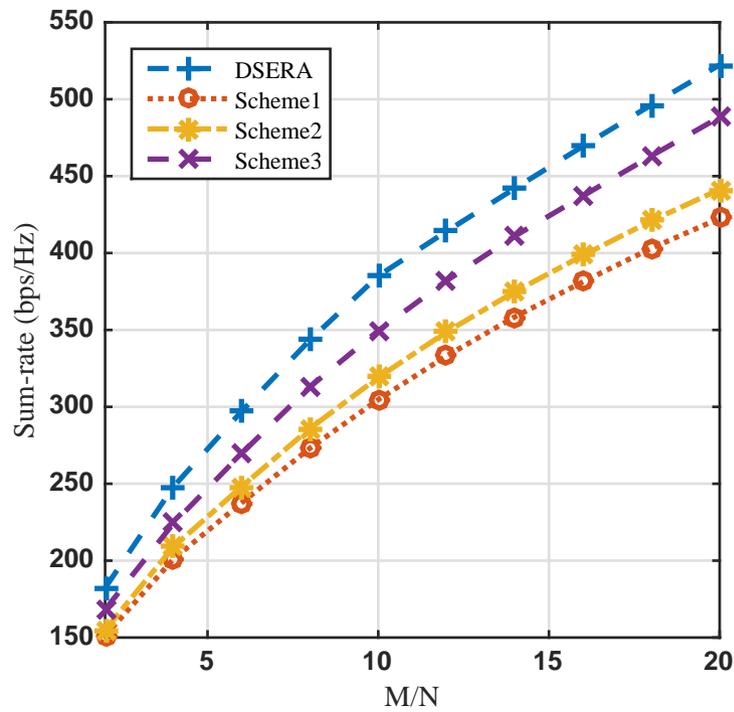}
    \caption{Sum-rate of different utility function definitions and price updating procedures versus the cell density when $N=10$.}
    \label{FigSumDistSchVsMN}
\end{figure}
\begin{figure}[!h]
    \centering
    \includegraphics[trim={0cm 0 0cm 0},clip,width=0.6\linewidth]{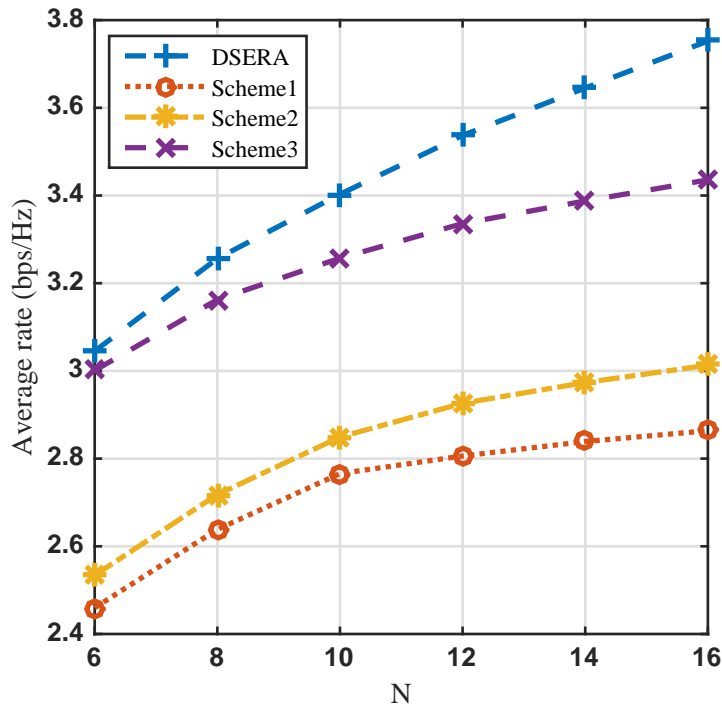}
    \caption{Average data rate of different utility function definitions and price updating procedures versus the number of CLs when $M/N=20$.}
    \label{FigAvgDistSchVsN}
\end{figure}
\begin{figure}[!h]
    \centering
    \includegraphics[trim={0cm 0 0cm 0},clip,width=0.6\linewidth]{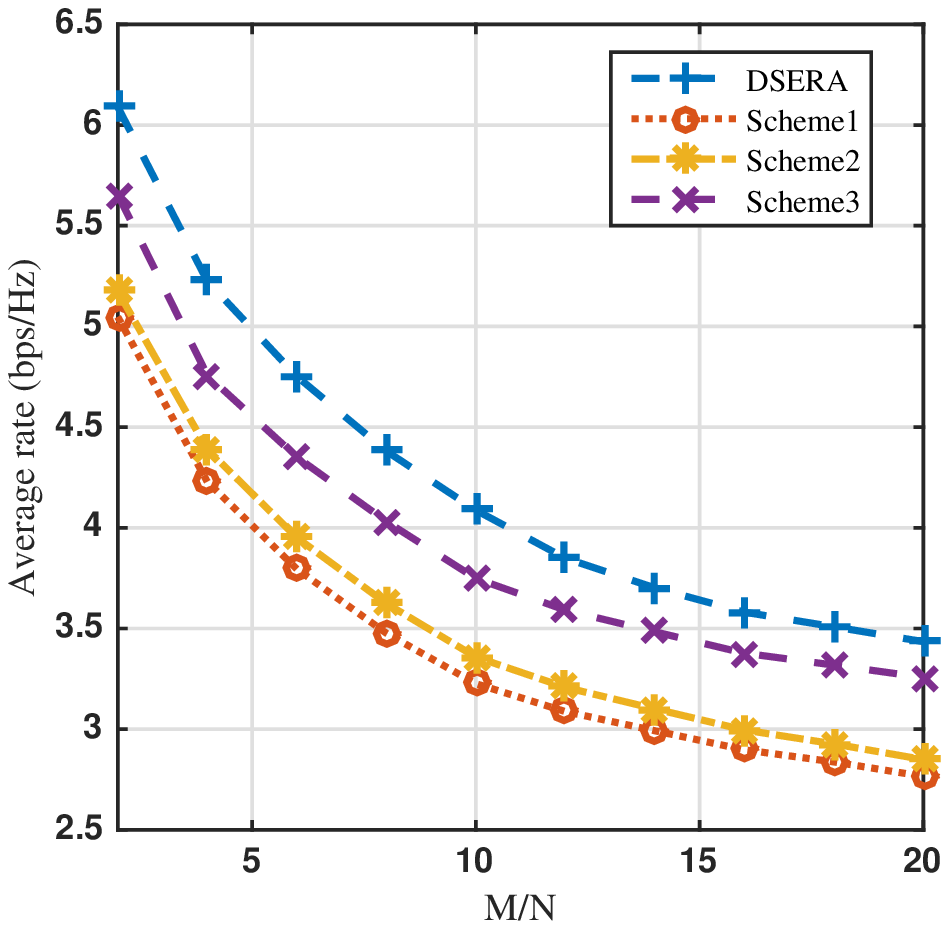}
    \caption{Average data rate of different utility function definitions and price updating procedures versus the cell density when $N=10$.}
    \label{FigAvgDistSchVsMN}
\end{figure}

In order to evaluate the performance and demonstrate the effectiveness of our proposed utility function and price updating procedure, we introduce other resource allocation algorithms with different utility functions and price updating procedures to be compared to our proposed resource allocation algorithm in the following:
\begin{itemize}
\item
\textbf{Scheme 1}: This algorithm is the modified version of the proposed resource allocation algorithm of 
Reference \cite{R011} 
where our innovative D2D pair admission procedure is added to it 
and the utility function of $d_j$ is formulated as
\begin{equation}
\label{UtilitySchemeOne}
U^d_j(\bm{\theta},\bm{P},\bm{\Psi})
= - p^d_j 
\sum_{i=1}^{N} 
\theta_i \psi_{i,j}
 \frac{h^{d,b}_j}
{h^c_i}.
\end{equation}
This algorithm can also be considered as a modified version of our proposed algorithm 
when no reward term is considered in the utility function definition, 
a limited number of interference prices are utilized, 
all the prices are updated at once, 
and no CUE utility function is defined.
\item
\textbf{Scheme 2}: 
By defining interference prices for all the interference links, this algorithm can be considered as the updated version of the Scheme 1 algorithm when multiple interference prices are utilized and the utility function of $d_j$ is expressed as
\begin{equation}
\label{UtilitySchemeTwo}
U^d_j(\bm{\theta},\bm{P},\bm{\Psi})= - p^d_j \Big(
\sum_{i=1}^{N} \psi_{i,j} \theta^{d}_{j} \frac{h^{d,b}_j}{h^c_i} + 
\sum_{i=1}^{N} \sum\limits_{k=1 \atop k\neq j}^M \psi_{i,j} \psi_{i,k} 
\theta_{j,k}^{d,d} \frac{ h^{d,d}_{j,k} }{ h^d_k } \Big).
\end{equation}
The Scheme 2 algorithm can be considered as a modified version of our proposed DSERA algorithm with no utility function reward term, no CUE utility function definition, and the whole price updating scheme.
\item
\textbf{Scheme 3}: This algorithm is the modified version of our proposed DSERA algorithm where the whole price updating scheme is utilized and the other definitions and steps remain unchanged.
\end{itemize}

Fig. \ref{FigSumDistSchVsN} demonstrates the sum-rate of different utility function definitions and price updating procedures for different numbers of CLs when $M/N=20$.
It is shown that the sum-rate of the Scheme 1 algorithm is lower than that of the other algorithms due to its interference-dependent utility function definition.
Considering the utility function of the Scheme 1 algorithm which is not rate-dependent, the data rate of CUEs and D2D pairs cannot be maximized but the minimum QoS requirements of CUEs and admitted D2D pairs can be guaranteed in a distributed manner.
Therefore, this algorithm has a lower sum-rate compared to the other ones.
By introducing prices for all the interference links, the Scheme 2 algorithm can result in a higher overall sum-rate compared to the Scheme 1 algorithm due to its better interference management.
The Scheme 3 and the Scheme 4 algorithms consider the data rate as a reward term in their utility functions which is the reason for the higher sum-rates of these two algorithms compared to the Scheme 1 and the Scheme 2 algorithms.
The Scheme 3 algorithm with the whole price updating scheme has a lower sum-rate  compared to the Scheme 4 algorithm which utilizes the step-by-step price updating scheme.
Since the sum-rate maximization is solved in a distributed manner by adjusting the price values, the step-by-step price updating scheme results in a higher sum-rate compared to the whole price updating scheme since a part of the prices is updated at each evaluation of phase two of the proposed algorithm.

Fig. \ref{FigSumDistSchVsMN} illustrates the sum-rate of the different utility function definitions and price updating procedures for different $M/N$ values when there are $N=10$ CLs.
The $M/N$ parameter which is the ratio of the number of D2D pairs to the number of CUEs can be considered as the cell density parameter.
It is shown that the gap between the sum-rate performances of different utility function definitions is small for low-density deployments while the gap is enhanced when the deployment density is increased.
Therefore, it can be concluded that the proposed utility function and the proposed price updating procedure of the DSERA algorithm result in a higher sum-rate and SE compared to the other utility function definitions and price updating schemes for different numbers of CLs and cell densities. 

The effect of utility function definition and price updating procedure on the average data rate is shown in Figs. \ref{FigAvgDistSchVsN} and \ref{FigAvgDistSchVsMN}. 
It should be noted that the overall sum-rate of the cell is the objective function of our optimization problem and the average data rate is not a performance evaluation metric.
From Fig. \ref{FigAvgDistSchVsN}, it can be concluded that the average data rates of the introduced algorithms increase with respect to the number of CLs.
Fig. \ref{FigAvgDistSchVsMN} shows the decreasing behavior of average data rates of the introduced algorithms with respect to the cell density.
It is verified that the DSERA algorithm is more effective for resource allocation in terms of average data rate compared to the other methods with different utility function definitions and price updating procedures when the number of CLs or the cell density is increased.

\subsection{Number of CLs}

\begin{figure}[!h]
    \centering
    \includegraphics[trim={0cm 0 0cm 0},clip,width=0.6\linewidth]{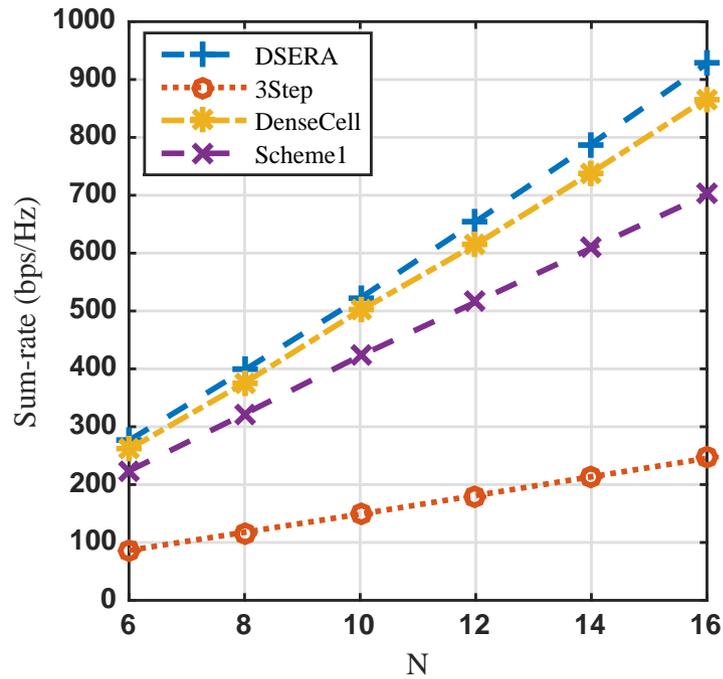}
    \caption{Sum-rate of different resource allocation methods versus the number of CLs when $M/N=20$.}
    \label{FigSumDistAllVsN}
\end{figure}
\begin{figure}[!h]
    \centering
    \includegraphics[trim={0cm 0 0cm 0},clip,width=0.6\linewidth]{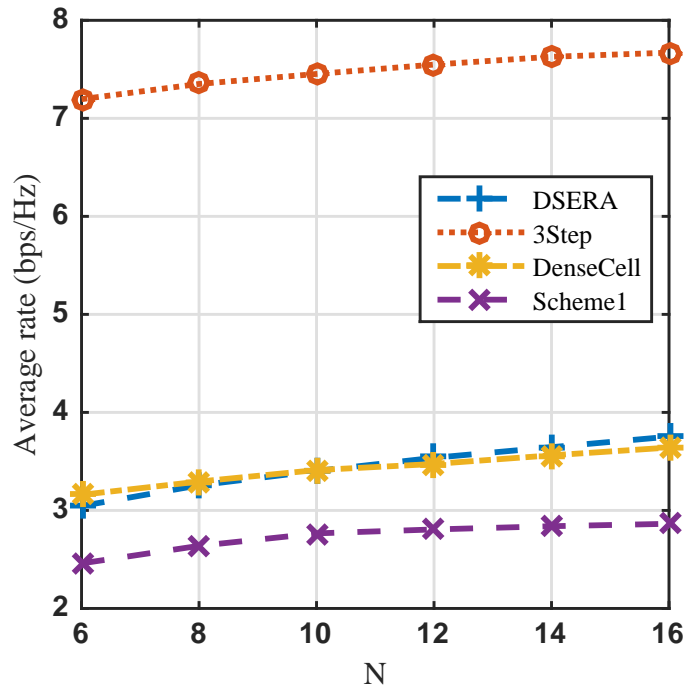}
    \caption{Average data rate of different resource allocation methods versus the number of CLs when $M/N=20$.}
    \label{FigAvgDistAllVsN}
\end{figure}

In order to compare our proposed resource allocation algorithm with other existing methods, the 3Step and the DenseCell resource allocation algorithms which are the slightly modified and improved versions of the resource allocation methods of Reference \cite{R003} and Reference \cite{R042_026}, respectively, are introduced in the following:
\begin{itemize}
\item
\textbf{3Step}:
In Reference \cite{R003}, it is assumed that the number of CUEs is equal to the number of D2D pairs and the Kuhn-Munkres \cite{R003_016} algorithm is used to find the optimal matching between CUEs and D2D pairs among $N^2$ possible matchings.
Due to the dense deployment of D2D pairs assumption, the number of D2D pairs is much larger than the number of CUEs.
Hence, we modify the Kuhn-Munkres algorithm to check all the $MN$ possible matchings.
The modified resource allocation algorithm is denoted as the 3Step algorithm.
\item
\textbf{DenseCell}:
In Reference \cite{R042_026}, a greedy resource allocation algorithm for D2D communications when $M>N$ has been proposed where the sum-rate of D2D pairs and the minimum QoS requirements of CUEs are considered.
The DenseCell method is the modified version of the proposed algorithm of Reference \cite{R042_026} where sum-rate and minimum QoS requirements of CUEs and admitted D2D pairs are considered.
\end{itemize}
It should be noted that the Scheme 1 algorithm is the improved version of the proposed algorithm of Reference \cite{R011} which has been introduced in the previous section.

Fig. \ref{FigSumDistAllVsN} shows the sum-rate of the DSERA algorithm compared to the other resource allocation algorithms for different numbers of CLs when $M/N=20$.
The Scheme 1 algorithm has a lower sum-rate compared to our proposed resource allocation algorithm due to the minimum QoS requirement guaranteeing behavior of this method as it has been explained in the previous section.
The 3Step algorithm assigns at most one D2D pair to each CL in an optimal manner and does not take advantage of the spatial reuse gain of D2D pairs.
Hence, the sum-rate performance of the 3Step algorithm is much less than the other methods.
The sum-rate performance of the DenseCell algorithm is slightly lower than the sum-rate performance of the DSERA algorithm but it has a computationally complex subchannel sharing protocol.

Fig. \ref{FigAvgDistAllVsN} demonstrates the average data rate of the DSERA algorithm compared to other existing methods for different numbers of CLs when $M/N=20$.
The average data rate of the Scheme 1 algorithm is lower than the other methods since the rate of CUEs and D2D pairs are not maximized in this algorithm but the minimum QoS requirements are just guaranteed.
The average data rate of the 3Step method is higher compared to the other ones since at most one D2D pair can reuse each CL.
Therefore, no interference exists among different D2D pairs and a higher average data rate can be achieved by admitted D2D pairs.
It should be noted that the average data rate behavior of all methods is increasing with respect to the number of CLs.

\subsection{Cell Density}

\begin{figure}[!h]
    \centering
    \includegraphics[trim={0cm 0 0cm 0},clip,width=0.6\linewidth]{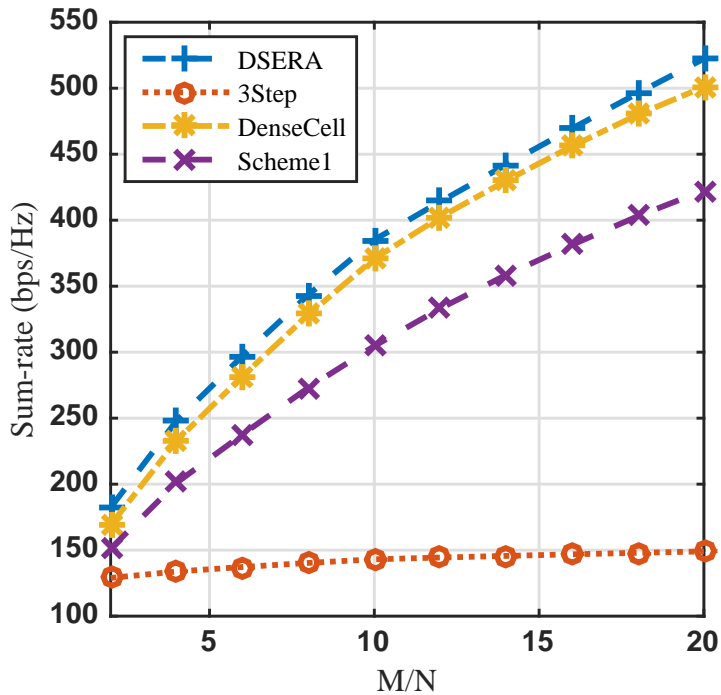}
    \caption{Sum-rate of different resource allocation methods versus the cell density when $N=10$.}
    \label{FigSumDistAllVsMN}
\end{figure}

Our proposed algorithm can also be compared to the other existing methods for different cell densities.
Fig. \ref{FigSumDistAllVsMN} compares the sum-rate of the DSERA algorithm with the sum-rate of the other methods for different cell densities when $N=10$ CLs are considered.
It is verified that the low complexity DSERA algorithm has higher sum-rate compared to the distributed and low complexity Scheme 1 algorithm as well as the complex and centralized DenseCell algorithm.
The sum-rate performance of the 3Step algorithm is lower than the other algorithms since it is assumed that at most one D2D pair can reuse each CL.

\begin{figure}[!h]
    \centering
    \includegraphics[trim={0cm 0 0cm 0},clip,width=0.6\linewidth]{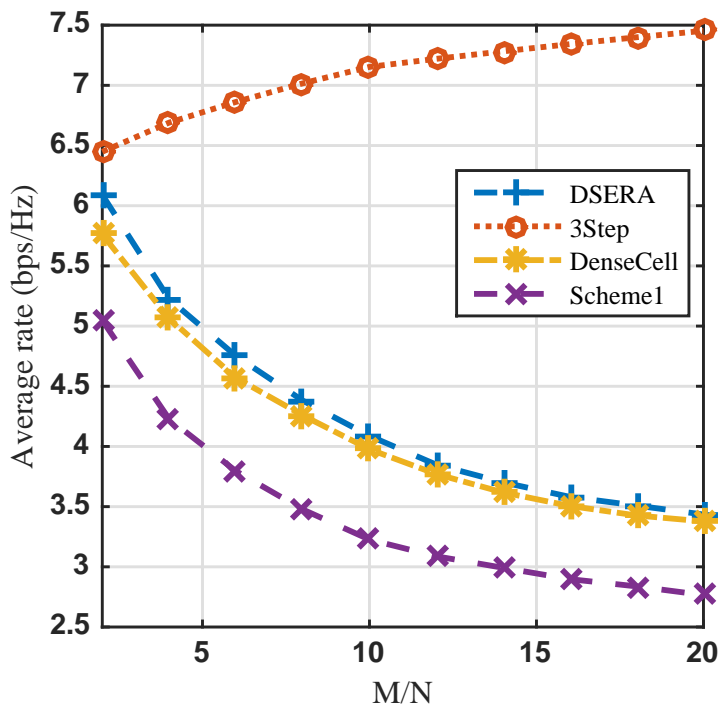}
    \caption{Average data rate of different resource allocation methods versus the cell density when $N=10$.}
    \label{FigAvgDistAllVsMN}
\end{figure}

Fig. \ref{FigAvgDistAllVsMN} compares the average data rate of the DSERA algorithm with the average data rate of the other methods for different cell densities when $N=10$ CLs are considered.
It is shown that the DSERA, the Scheme 1, and the DenseCell algorithms have a decreasing behavior with respect to the cell density since the cellular resources are allocated to multiple D2D pairs reusing each CL in these algorithms.
Considering the fact that at most one D2D pair can reuse each CL in the 3Step algorithm, each CL can be shared among a larger number of D2D pairs when the cell density is increased.
In other words, increasing the cell density can be considered as a diversity gain in the 3Step scheme.
Hence, the average data rate of the 3Step algorithm has an increasing behavior with respect to the cell density.

\section{Conclusion}
\label{ConcSec}
In this paper, a distributed framework of joint transmission power control and spectrum allocation for dense D2D communications underlaying cellular networks where multiple D2D pairs can reuse each CL and each D2D pair can reuse at most one CL is proposed to maximize the overall SE of the cell.
Our proposed low complexity distributed method (DSERA) maximizes the sum-rate of the cell using a novel pricing based resource allocation algorithm by transforming the mixed-integer non-convex centralized problem into a distributed algorithm consisting of two phases.
The sum-rate of the cell is maximized as a result of the first phase of the proposed algorithm where the derived closed-form solutions result in low complexity implementation of the proposed algorithm.
The QoS constraints of the CUEs and D2D pairs are guaranteed as a result of the price vector updating of the second phase. 
We proposed a fast and innovative admission control for D2D pairs which cause the proposed method to be practical for dense beyond 5G networks.
The distance-based cost term of the proposed novel utility function results in low signaling overhead and cause the proposed algorithm to be a practical one.
Furthermore, the SINR computation at each receiver using aggregate interference is discussed.
The numerical results are provided to verify the effectiveness of the proposed utility function definition and the price updating procedure in terms of overall sum-rate.
Additionally, the SE of DSERA is compared to other centralized and distributed resource allocation schemes to illustrate the higher sum-rate performance of DSERA.
It is demonstrated that the proposed method can effectively improve the SE of cellular networks while guaranteeing the minimum QoS requirements of CUEs and admitted D2D pairs in a distributed manner considering the dense deployment of D2D pairs in beyond 5G networks for different numbers of CLs and cell densities.

\section*{Data Availability}
Data sharing not applicable to this article as no datasets were generated or analysed during the current study.

\bibliography{Test}

\appendix
\section{Proof of equation (21)}
\label{AppA}
The optimization problem can be formulated as 
\begin{subequations}
\label{AppMaxEq}
\begin{alignat}{4}
&{p^c_i}^* = && \argmax_{p^c_i} ~~&&&& \log_{2} (1+p^c_i \Gamma^c_i) - p^c_i q^c_i,\\
& &&~~~~~\rm{s.t.}  &&&& 0 \leq p^c_i \leq P^{c,\rm{max}}_i.
\end{alignat}
\end{subequations}
The parameters $ \Gamma^c_i$ and $q^c_i$ are positive constants with respect to the optimization variable $p^c_i$.
The second derivative of the objective function can be computed as
\begin{equation}
\frac{d^2 f^c_i(p^c_i)}{{d p^c_i}^2} = -\frac{{\Gamma^c_i}^2}{(\ln 2)(1+p^c_i \Gamma^c_i)^2},
\end{equation}
which is always less than zero.
Therefore, the objective function is strictly concave with respect to $p^d_j$.

As the objective function is differentiable and strictly concave with respect to the only optimization variable, there exist only one point of zero slope.
The first derivative of the objective function can be expressed as
\begin{equation}
\frac{d f^c_i(p^c_i)}{{d p^c_i}} = \frac{\Gamma^c_i}{(\ln 2)(1+p^c_i \Gamma^c_i)}-q^c_i
\end{equation}
and the point of zero slope $\bar{p^c_i}$ can be computed as
\begin{equation}
\frac{d f^c_i(\bar{p^c_i})}{{d p^c_i}}=0 \Rightarrow \bar{p^c_i}=\frac{1}{(\ln 2) q^c_i} - \frac{1}{\Gamma^c_i}.
\end{equation}
The point $p^c_i=\bar{p^c_i}$ is the global maximum of the unconstrainted form of  \eqref{AppMaxEq}.
Hence, the optimization problem \eqref{AppMaxEq} is solved if the point $p^c_i=\bar{p^c_i}$ is inside the feasible region which can be expressed as
\begin{equation}
\label{AppEq1}
{p^c_i}^* =\bar{p^c_i}=\frac{1}{(\ln 2) q^c_i} - \frac{1}{\Gamma^c_i}, ~ \textrm{if}~~ 0 \leq \bar{p^c_i} \leq P^{c,\rm{max}}_i.
\end{equation}
If the point $p=\bar{p^c_i}$ is outside the feasible region and $\bar{p^c_i} <0$, it can be concluded that 
\begin{equation}
\frac{d f^c_i(p^c_i)}{{d p^c_i}} < 0, ~ ~\textrm{if}~~0 \leq p^c_i \leq P^{c,\rm{max}}_i,
\end{equation}
due to the fact that $\frac{d f^c_i(\bar{p^c_i})}{{d p^c_i}} = 0$,  $\bar{p^c_i} <0$, and $\frac{d^2 f^c_i(p^c_i)}{{d p^c_i}^2} < 0, ~\forall p^c_i$.
Therefore, the point of maximum value for the objective function is $p^c_i=0$ which can be expressed as
\begin{equation}
\label{AppEq2}
{p^c_i}^* =0, ~ \textrm{if}~~ \bar{p^c_i} \leq 0.
\end{equation}
If the point $p=\bar{p^c_i}$ is outside the feasible region and $\bar{p^c_i} >0$, it can be concluded that 
\begin{equation}
\frac{d f^c_i(p^c_i)}{{d p^c_i}} > 0, ~ ~\textrm{if}~~0 \leq p^c_i \leq P^{c,\rm{max}}_i,
\end{equation}
due to the fact that $\frac{d f^c_i(\bar{p^c_i})}{{d p^c_i}} = 0$,  $\bar{p^c_i} >0$, and $\frac{d^2 f^c_i(p^c_i)}{{d p^c_i}^2} < 0, ~\forall p^c_i$.
Therefore, the point of maximum value for the objective function is $p^c_i=P^{c,\rm{max}}_i$ which can be expressed as
\begin{equation}
\label{AppEq3}
{p^c_i}^* =P^{c,\rm{max}}_i, ~ \textrm{if}~~ \bar{p^c_i} \geq P^{c,\rm{max}}_i.
\end{equation}
Considering \eqref{AppEq1}, \eqref{AppEq2}, and \eqref{AppEq3} the final solution can be expressed as
\begin{equation}
{p^c_i}^* = \mathcal{F}(q^c_i,\Gamma^c_i,P^{c,\rm{max}}_i)=
\left\{
	\begin{array}{ll}
		0, 
		&\mbox{if}~ (\frac{1}{\ln(2)~q^c_i} - \frac{1}{\Gamma^c_i}) \leq 0 \\
		\frac{1}{\ln(2)~q^c_i} - \frac{1}{\Gamma^c_i}, ~
		&\mbox{if}~ 0 < (\frac{1}{\ln(2)~q^c_i} - \frac{1}{\Gamma^c_i}) < P^{c,\rm{max}}_i \\
		P^{c,\rm{max}}_i, 
		&\mbox{if}~ (\frac{1}{\ln(2)~q^c_i} - \frac{1}{\Gamma^c_i}) \leq 0
	\end{array}
\right.,
\end{equation}
and the proof is complete.

\end{document}